\documentclass{mn2e}
\usepackage{psfig}

\title[Alignment of disc and stellar rotation axes]
{Chaotic star formation and the alignment of 
  stellar rotation with disc and planetary orbital axes}

\author[M. R. Bate, G. Lodato \& J. E. Pringle] {
  M. R. Bate$^1$,
  G. Lodato$^{2,3}$ and
  J. E. Pringle$^{2,4}$ \\
  $^1$School of Physics, University of Exeter, Stocker Road, Exeter, EX4 4QL\\ 
  $^2$Theoretical Astrophysics Group, University of Leicester, Leicester LE1 7RH\\ 
  $^3$Dipartimento di Fisica, Universit\'a di Milano, Via Celoria 16, I-20133, Milano, Italy\\ 
  $^4$Institute of Astronomy, University of Cambridge, Madingley Road, Cambridge CB3 0HA}

\date{Accepted for publication in MNRAS}
\volume{000}

\pagerange{\pageref{firstpage}--\pageref{lastpage}} \pubyear{2009}

\begin{document}

\label{firstpage}

\maketitle

\begin{abstract}
We investigate the evolution of the relative angle between the stellar rotation axis
and the circumstellar disc axis of a star that forms in a stellar cluster from the collapse
of a turbulent molecular cloud.  This is an inherently chaotic environment with
variable accretion, both in terms of rate and the angular momentum of the material, and dynamical 
interactions between stars.  We find that the final stellar rotation axis and disc spin
axis can be strongly misaligned, but this occurs primarily when the disc is truncated
by a dynamical encounter so that the final disc rotation axis depends simply on what
fell in last.  This may lead to planetary systems with orbits that are misaligned with the
stellar rotation axis, but only if the final disc contains enough mass to form planets.
We also investigate the time variability of the inner disc spin axis, which is likely to
determine the direction of a protostellar jet.  We find that the jet direction varies more strongly for
lighter discs, such as those that have been truncated by dynamical interactions or
have suffered a period of rapid accretion.  Finally, we note that variability of
the angular momentum of the material accreting by a star implies that the
internal velocity field of such stars may be more complicated than that of aligned differential
rotation.
\end{abstract}

\begin{keywords}
  accretion, accretion discs -- stars: formation -- stars: interiors -- stars: rotation -- planetary systems: formation -- planetary systems: protoplanetary discs
\end{keywords}

\section{Introduction}

Over the last 20 years or so our picture of the environment in which
stars form and the way in which they do so has changed
considerably. With a few notable exceptions (for example, Larson
1978), prior to 1990 the process of star formation was treated by
forming one star at a time (see, for example the reviews by Shu, Adams
\& Lizano, 1987; McKee \& Ostriker, 2007). The formation of binary
stars was also considered, but mainly through numerical simulation of
the dynamics of rotating rings or bars of material (see reviews by
Bodenheimer \& Burkert, 2001; Bonnell, 2001). In all these scenarios,
the symmetries are such that the rotation axes of any stars formed
tend to remain aligned with the rotation axis of any surviving
circumstellar material or disc. Thus these simple models all predict
that protostellar discs, and therefore any planetary sytem formed
within them, will be closely aligned with the rotation axis of the
central star. This expectation is reinforced by observation of the one
system for which we have good data -- the Solar System (Tremaine,
1991). We need to be aware, however, that while the solar system may
be a good guide to the formation of planetary systems in general,
because of our special place within it, it may also be atypical in
some ways (e.g. Beer et al., 2004). In this paper we consider how an
examination of the degree to which stellar rotation axes are aligned
with their protostellar discs and/or planetary systems can give us
information about the way in which stars form. This is becoming a
relevant consideration, given a recent amount of growing evidence of
star-disc misalignment from observations of spin-orbit misalignment of
transiting exoplanets (H\'ebrard et al., 2008;  Winn et al. 2009a, 2009b; 
Pont et al., 2009a, 2009b; Gillon 2009; Johnson et al., 2009;
Narita et al. 2009).

The fact that many, if not most, stars are members of binary or
multiple systems has led to the argument that the star formation
process must be much more dynamical and interactive than the simple
pictures described above (e.g. Pringle, 1989, 1991; Clarke \& Pringle,
1991). Early multiplicity surveys of pre-main-sequence stars in nearby
star-forming regions confirmed that most low-mass stars in the solar
neighbourhood form in multiple systems (e.g. Leinert et al., 1993;
Ghez, Neugebauer \& Matthews 1993; Simon et al., 1995). Indeed 
most stars within the Galaxy appear to form
in clustered environments (see the reviews by Lada \& Lada, 1995,
Clarke, Bonnell \& Hillenbrand, 2000) and there is evidence that disc
sizes and presumably disc lifetimes are reduced in binary as compared
to single stars (Cieza et al., 2009). The theoretical expectation is
that close encounters in the early stages of formation when stars have
substantial discs lead to substantial disc truncation and moreover,
that the tidal effects of such encounters can also lead to the outer
parts of the disc becoming strongly misaligned or warped (Heller,
1995; Hall, Clarke \& Pringle, 1996; Larwood, 1997; Pfalzner, 2004;
Moeckel \& Bally, 2006). Such interactions imply that there is a
strong possibility that the stellar rotation axis (determined by the
angular momentum of material accreted early which forms the stars) is
not aligned with the disc rotation axis (which might correspond to the
angular momentum of material accreted later which forms the disc, and
eventually planets).

Recent numerical simulations of the star formation process (Bate,
Bonnell \& Bromm, 2002a,b, 2003; Bate \& Bonnell 2005; 
Bate 2005, 2009a,b) have
considered the formation of stars within clusters, starting
with initial turbulent conditions for the gas which are designed to
mimic those seen within star forming molecular clouds. The picture
which results involves turbulent and chaotic motions of both gas and
stars, with disc fragmentation, competitive accretion and close
dynamical interactions all playing a role. Close star-disc encounters
occur throughout the star formation process at a variety of
orientations. From this perspective it might seem reasonable to
conclude the star-disc misaligment  would be the norm rather than the
exception. Here we investigate this possibility.

Because of the large range of scales involved it is not feasible to
follow the evolution of gas starting from the turbulent molecular
cloud (tens of parsecs) down the the sizes of protostellar discs (tens
of au) and further to the accretion of matter onto the protostars
themselves (a few solar radii). Numerical simulations of the star
formation process begin with a turbulent molecular cloud but truncate
the computation typically at a resolution scale of several au by means
of sink particles. In this paper we are interested in what happens to
the gas when it falls within the sink particles. We are therefore
constrained to carry out a highly idealised computation using a
strongly simplified, and not necessarily self-consistent, set of
assumptions. We consider the evolution gas falling into a sink
particle taken from a particular simulation. In Section 2 we describe
the simulation. In Section 3 we describe the
simplifications we have made in order to model the evolution of the
gas when it has fallen into the sink particle from the formation of
the sink particle ($\approx 10^{-3}~{\rm M}_\odot$) until after accretion
onto the sink particle has finished and the 'star' is more or less
fully formed with mass $\approx 0.2~{\rm M}_\odot$. In Section 4 we present
the results of our calculations. We discuss the implications in
Section 5.

\section{Simulation data}

We are interested here in illustrating the growth of a particular
protostellar core. The data we use comes from a computation by Bate,
Bonnell \& Bromm (2003) which describes the formation of a
cluster of stars from a turbulent self-gravitating parcel of dense
(molecular) gas. The details of the numerical procedures are described
fully in that paper, and here we draw attention to those details of
particular relevance to the calculations below. 

The calculation was performed using a 3D smoothed particle
hydrodynamics (SPH) code. The smoothing lengths are variable in time
and space, subject to the constraint that the number of neighbours for
each particle must remain approximately constant at $N_{\rm neigh} =
50$. The initial cloud is spherical and uniform in density with a mass
of 50~${\rm M}_\odot$ and diameter of 0.375 pc (77,400 au). At the initial
temperature of 10 K, the thermal Jeans mass is 1~${\rm M}_\odot$. The free
fall timescale of the cloud is $t_{\rm ff} = 1.90 \times 10^5$ yr.  An
initial suspersonic turbulent velocity field is imposed on the cloud
with rms Mach number ${\cal{M}} = 6.4$. The field is divergence-free
and random Gaussian with a power spectrum $P(k) \propto k^{-4}$, where
$k$ is the wavenumber. The local Jeans mass is resolved throughout the
calculation. The number of particles used is $3.5 \times 10^6$, which
means that each particle has a mass of $1.43 \times 10^{-5}~{\rm M}_\odot$.

The calculation uses a barotropic equation of state that is isothermal
up to densities of $10^{-13}$ g~cm$^{-3}$, but with the temperature of
the gas increasing at higher densities.  In this way, the 
calculations mimic the opacity limit for fragmentation 
(Low \& Lynden-Bell 1976; Rees 1976).
Despite this, because of computational constraints, it is necessary 
to excise the high density protostellar cores from the bulk of the
computation. This is done by inserting a sink particle (Bate et al.,
1995) whenever the central density of a pressure-supported fragment
exceeds a certain density, $\rho_{\rm crit} = 10^{-11}$ g
cm$^{-3}$. The sink particle is formed by replacing the SPH gas
particles contained within a radius $R_{\rm sink} = 5$ au of the
densest gas particle by a point mass with the same mass and
momentum. Any gas that later falls within this radius is accreted by
the point mass if both (a) it is bound and (b) its specific angular
momentum is less than that required to form a circular orbit at radius
$R_{\rm sink}$ from the sink particle. Sink particles interact with the
gas only via gravity and by accretion. Sink particles interact with
each other only by gravity. The gravitational acceleration between two
sink particles is softened within a distance of 4 au. Thus sink
particles cannot merge with each other. The typical initial mass of a
sink particle in this calculation is $\approx 10^{-3}~{\rm M}_\odot$.

\section{Evolution Equations}

As we have seen, because of resolution constraints, in the numerical
simulations of chaotic star formation the 'stars' formed are
represented by sink particles. Because of this, from the simulations
we only have very basic information about material being accreted onto
the protostellar cores. We have information about the rates of
accretion of both mass and angular momentum onto the sink particles,
but note that the accretion is discretized into that of individual SPH
particle masses of $\approx 10^{-5}~{\rm M}_\odot$. The sink particle radius
is set at 5 au., which is slightly less than the peak in the
distribution of binary star separations for low mass stars. We also
have information about the distance to the nearest neighbouring sink
particle as a function of time. It is usually the case that for some
of the time this distance is less than the radius of the sink
particles.

Since the information about the accretion process onto sink particles
is so limited, we are strongly constrained in how we are able to model
what might happen to the gas when it is accreted within the sink
particle radius of 5 au. We proceed by making the simplest assumptions
we can about the gas flow within the sink particle, noting that some
of the implications of these are of necessity at times not
self-consistent.

Within the sink particle, we model the gas flow as a twisted accretion
disc around a central mass.  The evolution of such a disc is computed
in the manner described by Pringle (1992). The disc surface density is
given as a function of time and radius by $\Sigma(r,t)$. Each disc
annulus, at radius $R$, is assumed to rotate with angular velocity
$\Omega(r,t)$ around the central mass, and to have a spin in the
direction of the unit vector ${\bf l}(R,t)$. Thus the disc locally has
an angular momentum density given by ${\bf L}(R,t) = \Sigma R^2 \Omega
{\bf l}$. We shall also assume (cf. Lin \& Pringle 1990) that any
matter being added to the disc is added at the radius corresponding to
its specific angular momentum, so that it arrives already on a
circular orbit and in centrifugal balance.

Then (Pringle 1992) the evolution of the angular momentum density is
given by
\begin{eqnarray}
\label{Levolution}
\frac{\partial {\bf L}}{\partial t} & = & \frac{1}{R}
\frac{\partial}{\partial R} \left\{ \frac{(\partial/\partial R)[\nu_1
    \Sigma R^3 (-\Omega^\prime)]}{\Sigma (\partial/\partial R)(R^2
    \Omega)} {\bf L} \right\} \nonumber \\
& & + \frac{1}{R} \frac{\partial}{\partial R} \left[ \frac{1}{2} \nu_2 R
  |{\bf L}| \frac{\partial {\bf l}}{\partial R} \right] \nonumber \\
&  & + \frac{1}{R} \frac{\partial}{\partial R} \left\{ \left[ \frac{(1/2)
      \nu_2 R^3 \Omega | \partial {\bf l}/\partial R
      |^2}{(\partial/\partial R)(R^2 \Omega) } + \nu_1 \left( \frac{R
        \Omega^\prime}{\Omega} \right) \right] {\bf L} \right\}
\nonumber \\ & & + {\bf S}(R,t).
\end{eqnarray}
Here $\Omega^\prime = d \Omega/dR$, $\nu_1$ is the shear viscosity
normally associated with accretion discs and $\nu_2$ is associated
with straightening out the out-of-plane motions (i.e. the warp or
twist). The source term ${\bf S}(R,t)$ allows for the addition of
material.

When the sink particle forms it has a mass, typically around $\approx
10^{-3}~{\rm M}_\odot$, which is much less than the eventual mass of the
star. Thus during the accretion process we expect the overall mass of
the star+disc system to grow by a factor of around $10^2 - 10^3$. To
calculate the angular velocity $\Omega(R)$ we need to take account of
the central time-varying force contributions from both the central
star, mass $M_\ast(t)$ and the disc itself. Although an exact determination 
of $\Omega(R)$ could be made (e.g. Bertin \& Lodato 1999), 
in general we shall find
that the disc mass is much less than of the central star. Thus for
computational convenience (cf. Lin \& Pringle 1990) we adopt the
approximation
\begin{equation}
\label{angvel}
\Omega(R,t) = \left( \frac{G M(R,t)}{R^3} \right)^{1/2},
\end{equation}
where
\begin{equation}
M(R,t) = M_\ast(t) + \int_0^R \Sigma(R,t) \, 2 \pi R \, dR.
\end{equation}

In order to estimate the magnitudes of the viscosity terms it is
strictly necessary to compute the disc structure at each radius. For
protostellar discs this is not a straightforward matter and the
physical processes relevant at various radii are still a matter for
debate (see for example Lodato \& Rice 2004, 2005; 
Kratter, Matzner \& Krumholz, 2008; Terquem 2008).  
What would be required first is a detailed time-dependent model of the evolution
of the disc surface density distribution, similar to that carried out by
Lin \& Pringle (1990), but taking more recent concepts and understanding
into account (see, for example, Armitage, Livio \& Pringle 2001; 
Rice \& Armitage 2009; Zhu, Hartmann \& Gammie 2009). 
In addition, a secondary, but nevertheless important,
consideration would be to take account of how such discs might respond to
misalignments -- for example, the low viscosity regions (the dead zones)
might transfer warp in a wave-like, rather than in a diffusive, manner,
and gravitational bending torques might play a role in disc regions where
self-gravity is important (Papaloizou \& Lin 1995).

With these issues in mind, and in order to simplify matters in this
initial investigation, we make the following assumptions. We assume
that for each viscosity, i.e. for $i = 1,2$,
\begin{equation}
\nu_i = \alpha_i \left( \frac{H}{R} \right)^2 R^2 \Omega.
\end{equation}
Here $H$ is the disc semi-thickness, and we shall assume for
convenience that $H/R = 0.1$, independent of radius (cf. Bell et al.,
1997; Terquem, 2008).  Thus $\alpha_1$ is the usual Shakura \& Sunyaev
(1973) viscosity parameter and we take typically (cf. King, Pringle \&
Livio, 2007)
\begin{equation}
\alpha_1 = 0.02,
\end{equation}
although we also consider one case with $\alpha_1 = 0.2$.
Typically $\alpha_2 \gg \alpha_1$ (Papaloizou \& Pringle 1983) and we
shall take (Lodato \& Pringle 2007)
\begin{equation}
\alpha_2 = 2.
\end{equation}

As far as the evolution of the system is concerned, what matters most
are the relative sizes of the various physical timescales. Thus to
zeroth order, the details of the viscous processes are less important
than the magnitude. This gives rise to the viscous timescale for mass
flow through the disc which is then given by
\begin{equation}
\label{nu1time}
t_{\nu_1} \approx 9000 \left(\frac{\alpha_1}{0.02}\right)^{-1}  \hspace{-3pt} \left(\frac{H}{0.1R}\right)^{-2}  \hspace{-3pt} \left(\frac{M}{{\rm M}_\odot}\right)^{-1/2} \hspace{-5pt} \left(\frac{R}{5~{\rm au}}\right)^{3/2} {\rm yr}.
\end{equation}
The timescale on which warp is propagated through the disc is less
than this and is
\begin{equation}
\label{nu2time}
t_{\nu_2} \approx 90 \left(\frac{\alpha_2}{2}\right)^{-1}  \hspace{-3pt} \left(\frac{H}{0.1R}\right)^{-2}  \hspace{-3pt} \left(\frac{M}{{\rm M}_\odot}\right)^{-1/2} \hspace{-5pt} \left(\frac{R}{5~{\rm au}}\right)^{3/2} {\rm yr}.
\end{equation}

\subsection{Numerical details}

We take a fixed radial grid with inner radius $R_{\rm in} = 10
R_\odot$. The choice of an outer radius of the grid presents us with a
consistency problem.  Some of the sink particles in the calculation
have resolved discs outside of the sink particle radius.  The interactions
between these discs and the discs which we model within the
sink particles is not accounted for.
This is problematic because the timescale governing the propagation 
of warps is much shorter than the viscous evolution timescale.
Therefore, the direction of the spin axis in the disc inner parts
is affected by that of the spin axis in the disc outer parts, which
may contain most of the mass.  However, even if the simulation of
Bate et al.\ were to be performed again, there is no way to
provide a continuous and consistent link between what is inside 
and what is outside the sink particle.  When sink particles were 
originally invented, even the construction of boundary conditions 
that tried to minimise the unphysical effects on pressure and
viscosity that were introduced outside the sink particle accretion 
radius proved extremely
difficult to implement (Bate 1995; Bate, Bonnell \& Price 1995).
There is certainly no way to link the propagation of warps across
the accretion radius.  Instead, for the calculations presented here,
we have chosen one of the many sink particles that does not have a
resolved disc outside of the sink particle radius.

However, even for those sink particles not surrounded by
resolved discs, much of the material accreted onto the sink
particle has an angular momentum which would put it on a circular
orbit at just inside the sink particle radius of 5 au. Thus we expect
the disc evolution to lead to disc expansion outside 5 au.  This is of
course not permitted in the original calculation. However, it would be
unrealistic to choose the outer disc radius at 5 au and so to simply
remove all disc mass which expands beyond 5 au as that leads to a
significant fraction of the material accreted onto the sink particle
not being accreted by the central star. Instead it would be lost back
to the computational domain. Thus whatever choice we make leads to an
inconsistency. As a compromise we choose the outer disc radius to be
$R_{\rm out}= 50$ au and note that we therefore allow the disc to
expand outside the radius of the sink particle $R_{\rm sink} = 5$ au.

We employ an explicit first-order finite difference scheme, as
detailed in Pringle (1992), to solve the time-evolution of
equation~\ref{Levolution}. We take 120 logarithmically spaced grid
points, Thus the ratio of radii of neighbouring grid points is 1.06 so
that $dR/R = 0.06$. The timestep is adjusted to satisfy the usual
numerical stability criteria. For most of the time it is set by the
diffusion time-scale across the innermost radial zone.

At each time step the following procedure is carried out.

1. Mass is added to the disc at the appropriate radius.  This is the
radius at which the specific angular momentum of the added matter
equals that of the local disc material. The largest radius at which
matter can be added corresponds to the the radius of the sink particle
$R_{\rm sink} = 5$ au which is a factor of 10 smaller than the outer
disc radius at all times. If the added material has sufficiently small
angular momentum that it lies within the inner disc radius $R_{\rm in}
= 10 R_\odot$ then the mass and angular momentum is added to the
central star. Because the mass of individual SPH particles are quite
massive compared to the mass of the disc we spread the accretion of
mass onto the disc in time by assuming that at each time $t$ the
accretion rate is a constant given by the mass of an SPH particle
divided by the time between its arrival and that of its predecessor. 

2. At this stage because mass has been added, and because disc
evolution has taken place (see step 3), the angular velocity profile
of the disc must be adjusted. First the angular velocity profile is
updated using equation~\ref{angvel}. Then the amount of material in
each grid zone is adjusted to restore centrifugal balance. To achieve
this we move material inwards in the manner described by Bath \&
Pringle (1981; see also Lin \& Pringle, 1990). This process conserves
angular momentum to machine accuracy, but conserves mass only to the
order of the numerical scheme.

3. The evolution of the disc angular momentum distribution is computed
from equation~\ref{Levolution}. The numerical scheme is written in
such a manner that it conserves angular momentum to machine accuracy,
but conserves mass only to the order of the scheme. At the inner edge
of the grid we apply a zero-torque boundary condition.  Mass and 
angular momentum are freely accreted onto the
star, and the stellar mass $M_\ast(t)$ and angular momentum ${\bf
  L}_\ast(t)$ are appropriately updated.  At the outer edge of the grid
we allow angular momentum (and mass) to flow freely outwards.

\subsection{Mass conservation}

We have noted that the although the various steps in the numerical
scheme conserve angular momentum to machine accuracy (except for the
small loss at the outer disc edge), they do not conserve mass. We show
in the Appendix that for Step 2 this leads at each timestep to a
systematic overestimate of the mass added by a factor of order
$(dR/R)^2$. Similar considerations apply to Step 3. 

\subsection{Interaction with other sink particles}
\label{truncation}

At each time we have the distance to the nearest sink particle $R_{\rm
  next}(t)$ and have noted that it is often the case the $R_{\rm next}
< R_{\rm sink}$. We are only able to take the interaction with the
neighbouring sink particle into account in an idealised fashion. To do
so we take note of the finding (e.g. Hall et al, 1996) that the main
effect of the interaction between a disc and the fly-by of a point
mass is to unbind the outer disc regions and so to truncate the
disc. Here we shall mimic this interaction by removing all disc
material which is at radii $R < 0.5 R_{\rm next}$.  We also note that
an encounter is likely to perturb the angular momentum of the
disc, particularly near the truncation radius.  We are forced to
neglect any such effect, but we note that a typical dynamical
encounter is also associated with the accretion of new material
into the sink particle and that the addition of this new material, 
whose angular momentum is likely to be uncorrelated with that 
of the existing star and disc, is likely to be more important 
than the tidal effect of the encounter on the pre-existing disc.
Indeed, as will be seen in the calculations presented below,
when truncation of the disc through interactions with other sink
particles is included, the stellar and disc axes ending up strongly 
misaligned because the angular momentum of the new material
is unrelated to the star's angular momentum.  If anything, including 
the perturbation suffered by the pre-existing disc during an encounter 
would typically serve only to increase the degree of misalignment, 
not decrease it, and therefore would not alter the fundamental result.

\subsection{Spin directions}

Since we know the total angular momentum accreted onto the central
star we can compute the direction of that vector in space. We use an
arbitrary fixed coordinate system and denote the space direction of
the stellar spin vector in spherical polar coordinates as
$(\theta_\ast(t), \phi_\ast(t))$. If and when the star is able to smooth
out any internal differential rotation this would represent the
stellar rotation axis.

We also measure the spin direction of the disc. We take this to be
represented by the spin direction $(\theta_{\rm d}(t), \phi_{\rm
  d}(t))$ of the annulus close to the inner disc edge at $\approx 3.2
R_{\rm in} = 32 R_\odot$. We choose this radius as it gives a
reasonable estimate of the direction in which any jet might be
launched, since the velocity of such jets indicates that they must be
launched from close to the central star (e.g. Pringle, 1993; Livio,
1997; Price, Pringle \& King 2003). In addition, by the end of the
calculation at a time of $5 \times 10^4$ yr, taken to be well after
the last accretion event, the disc is essentially coplanar, and so
this direction represents the plane of the protostellar disc and any
subsequent planetary system.

\section{Results}

\begin{figure}
\psfig{file=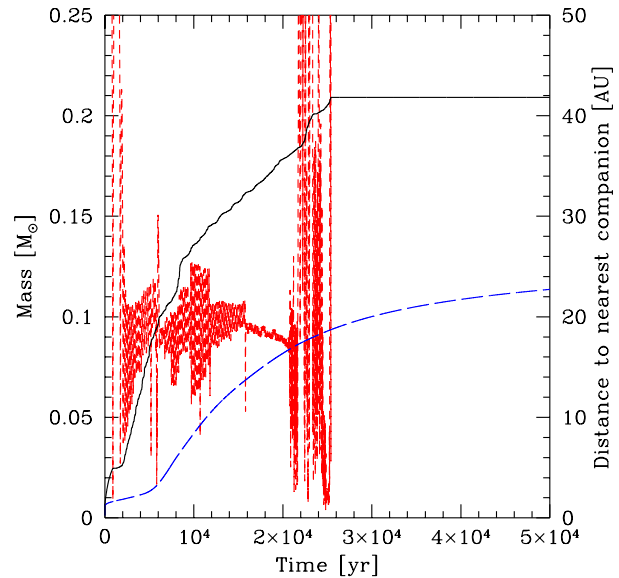,angle=-0,width=\hsize}
\caption{\label{N1}  Total mass accreted by the sink particle 
(solid black line), distance to the nearest other 
star in au (short-dashed red line), and mass accreted by
the central star (long-dashed blue line) as functions of time.
Although the nearest other star approaches within a few au,
the effect of the encounters are ignored for the evolution 
of the disc in this case.}
\end{figure}

In this section we consider the accretion history of one particular
sink particle taken from the 50 which formed
during the calculation.  Because of the inherent uncertainties and
inconsistencies of dealing with closely interacting sink particles we
choose to follow a sink particle which ends up as a single star. The
total mass accreted onto the sink particle is shown as a function of
time from the formation of the sink particle 
in Figure \ref{N1} (black solid line). It can be seen that accretion continues until a
time of $8 \times 10^{11}$ s $= 2.5 \times 10^4$ yr. The total mass
accreted is $4 \times 10^{32}$ g $\approx 0.2~{\rm M}_\odot$. Also in
Figure \ref{N1} we show the distance to the nearest sink particle $R_{\rm
  next}$ as a function of time (red short-dashed line). The sink particle
is the second sink particle that formed in the original hydrodynamical
calculation and starts its evolution as a companion to the first sink particle 
formed in the calculation.  As can be seen from Figure \ref{N1},
for some $2 \times 10^4$ yr these two sink particles form a binary with
separation of around 20 au. After that, the binary encounters a tighter binary
consisting of the 3rd and 10th objects to form in the calculation.  After a
chaotic interplay lasting about 3000 years, during which the distance of
closest approach occasionally become as small as 1 au, the sink particle
we are considering is ejected from the system (at the time of $2.5 \times 10^4$ yr
in Figure \ref{N1}) leaving objects 3 and 10
as a very tight 1 au binary (that survives until the end of the calculation)
with the first sink particle as a wide companion to the tight binary 
with a separation of about 60 AU.  This wider companion does not survive until the
end of the calculation -- it is unbound during another dynamical encounter  
5000 years later. Accretion onto the sink particle that we consider in this paper 
comes to an abrupt end when it is ejected from the multiple system and
is left as a single star. The computation
continues until a time $5 \times 10^4$ yr has elapsed.

\begin{figure}
\psfig{file=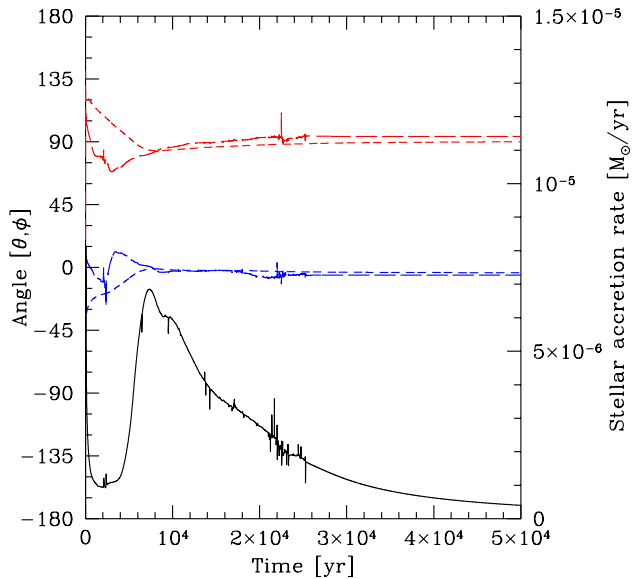,angle=-0,width=\hsize}
\caption{\label{N2}  The mass accretion rate of the central
star (solid black line) and the orientation angles of the 
stellar rotation axis and the rotation axis of the inner 
disc as functions of time.  The stellar rotation axis angles
($\theta_*=[0,180],\phi_*=[-180,180]$ degrees) are given by short-dashed lines in
red and blue, respectively.  The inner disc axis angles
$(\theta_{\rm d},\phi_{\rm d})$ and given by long-dashed
lines.  After the first 5000 years or so the stellar rotation
and inner disc axes are well aligned.  In this case, 
the effect of encounters with other stars on the disc
evolution is ignored.}
\end{figure}

\begin{figure}
\psfig{file=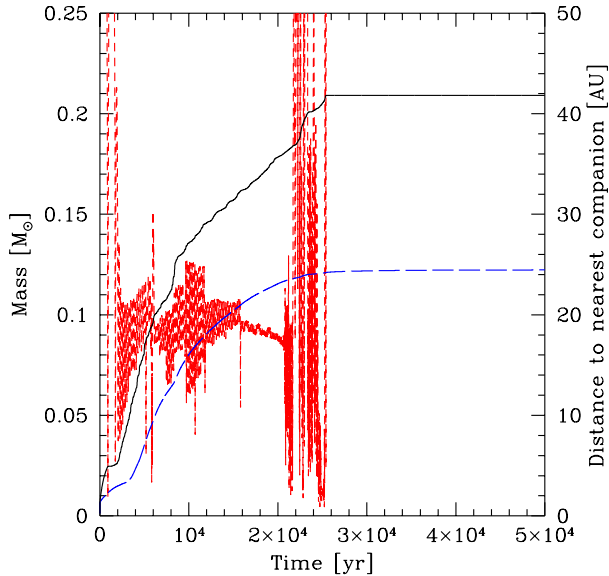,angle=-0,width=\hsize}
\caption{\label{N5}  Same as Figure \ref{N1}, except that
the disc viscosity is an order of magnitude larger at $\alpha_1=0.2$.  
Note that the mass passes through the disc more quickly than
in the low viscosity case and so the central star grows in mass more
quickly and because the disc drains quickly, accretion onto the
central star stops soon after the accretion into the sink particle ends.
The disc does not maintain a significant reservoir of material.}
\end{figure}

\begin{figure}
\psfig{file=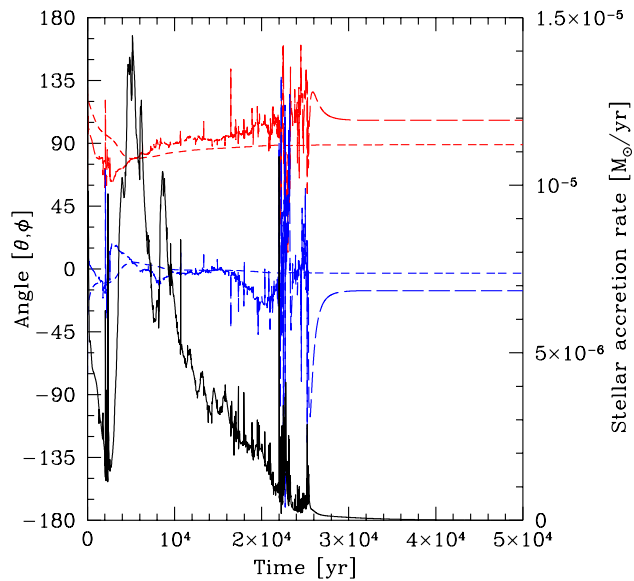,angle=-0,width=\hsize}
\caption{\label{N6}  Same as Figure \ref{N2}, except that
the disc viscosity is an order of magnitude larger at $\alpha_1=0.2$.
Note that the mass accretion rate on to the central
star is higher than the low viscosity case earlier, 
but the accretion onto the central star dies off soon after the
mass accretion into the sink particle ceases.  In addition, 
the stellar and inner disc axes are more misaligned than in
the low viscosity case.  Because accreted material passes quickly
through the disc the mass of the disc is low and its angular momentum
depends more sensitively on the material most recently accreted.}
\end{figure}

As can be seen from Figure \ref{N1} the accretion rate is not a
constant. Moreover, the accretion onto the sink particle occurs from a
variety of directions. We have seen that it is not possible to make a
set of fully consistent assumptions about how to treat the material
that falls into the sink particle. For this reason we shall consider
two extreme possibilities. First, we follow the evolution of the disc
ignoring any interaction with neighbouring sink particles. In this way
we model a star forming in the centre of a protostellar disc which is
being fed material from a variety of directions. Second, we take
account of the interaction of neighbouring sink particles which pass
through the disc by simply truncating the disc in the manner described
in Section~\ref{truncation} and removing the truncated matter. This
is, of course, not fully realistic as in reality the material should
be put back into the SPH computational domain. There it would interact
with other circumstellar material and would stand a chance of being
re-accreted.

\subsection{Without sink particle interaction}

We first consider the evolution of the system without taking account
of any interaction with the neighbouring sink particle. As we
mentioned above, although not strictly self-consistent, this serves to
illustrate the general behaviour of a star-disc system which is
accreting material with a wide range of angular momenta.

We consider the case when the disc viscosity is given by $\alpha_1 =
0.02$. In Figure \ref{N2} we show the accretion rate onto the central star
as a function of time (solid black line) and the time-evolution of
the directions of the stellar rotation axis($\theta_\ast, \phi_\ast$; short-dashed lines)
and of the rotation axis of the inner disc $(\theta_{\rm d}, \phi_{\rm
  d}$; long-dashed lines). At the end of the computation the stellar mass is $M_\ast =
0.113~{\rm M}_\odot$ and the disc mass is 
$M_{\rm d} = 3.4 \times 10^{-2}~{\rm M}_\odot = 0.3~M_\ast$
(self-gravity is likely to play a role in providing the effective viscosity of the
disc in this particular case; Lodato \& Rice 2004, 2005). 
Thus about a quarter of the mass accreted onto
the sink particle has been lost at the outer disc edge. This is to be
expected since much of the matter is accreted with an angular momentum
corresponding to a centrifugal radius $\approx R_{\rm sink}$ and
angular momentum conservation leads to the disc expanding beyond this
radius. The fraction of matter required to carry away the angular
momentum to radius $R_{\rm out}$ is approximately 
$\sqrt{R_{\rm sink}/R_{\rm out}} \approx 0.32$.

For $\alpha_1 = 0.02$ the accretion timescale (Equation~\ref{nu1time})
from where most the matter is accreted into the disc ($< 10^4$ yr at
radii $R < R_{\rm sink} = 5$ au) is comparable to the typical
timescale for the growth of stellar mass ($\approx 2 \times 10^4$
yr). Thus, there is still a non-negligible amount of mass in the disc
at the end of the computation. At that time the accretion rate onto
the star is $\dot{M} \approx 3.8 \times 10^{-7}~{\rm M}_\odot$
yr$^{-1}$. Comparing these values with the conventional picture of
the evolution of low-mass stars, our modelling corresponds to the
Class 0/I phases of protostellar evolution and ends at a stage comparable
to the beginning of the Class II or T-Tauri phase, both in terms of disc 
mass and stellar accretion rates.  
During the evolution, the stellar rotation axis and the disc spin axis both
converge quite quickly to their final values. Because the disc
evolution timescale is comparable to the stellar accretion timescale,
although its spin direction is somewhat influenced by late accretion
of material (especially when that material is accreted at radii close
to the inner disc radius where the disc spin is measured) it does not
differ substantially from that of the star. At the end of the
computation the angular distance between these two axes is
$4.2^\circ$.

\begin{figure}
\psfig{file=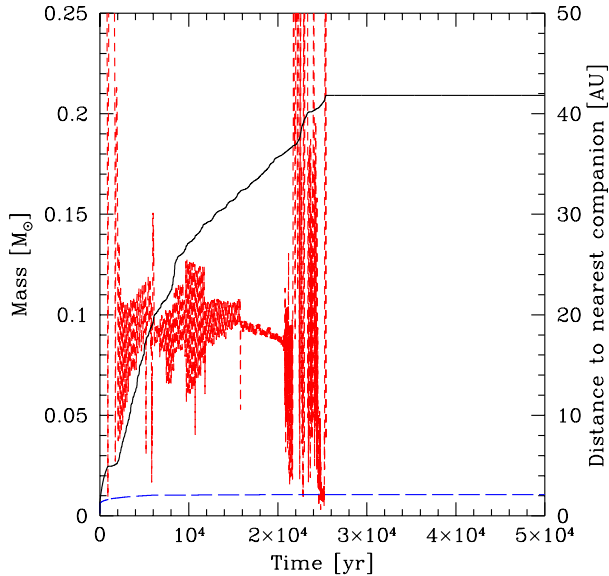,angle=-0,width=\hsize}
\caption{\label{N3}  Same as Figure \ref{N1}, except that
the disc is truncated by the encounters with the nearest
other star.  Note that the mass accreted by the central
star is dramatically reduced when the effects of the
encounters are taken into account because the disc is
repeatedly truncated.}
\end{figure}

\begin{figure}
\psfig{file=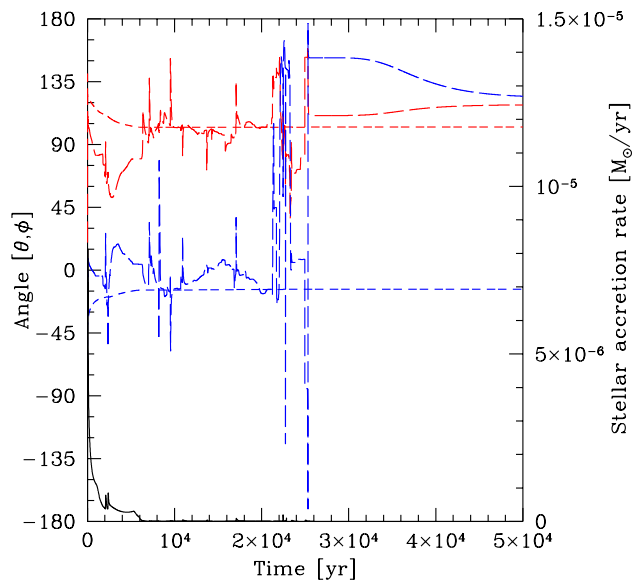,angle=-0,width=\hsize}
\caption{\label{N4}  Same as Figure \ref{N2}, except that
the disc is truncated by the encounters with the nearest
other star.  Note that the mass accretion rate on to the central
star is dramatically reduced when the effects of the
encounters are taken into account because the disc is
repeatedly truncated.  In addition, throughout most
of the evolution the stellar and inner disc axes are
significantly misaligned.  When the star is finally
ejected as a single object this misalignment persists.}
\end{figure}

In Figures \ref{N5} and \ref{N6} we consider an identical case, except
that the disc viscosity is increased by an order of magnitude to
$\alpha_1 = 0.2$.  In this case the accretion timescale is much less than
the timescale for the growth of the stellar mass.  Thus, the gas passes
through the disc quickly, the central star grows more quickly in mass
(Figure \ref{N5}),
the disc mass is lower at any given time, and significant accretion from
the disc onto the central star ceases at essentially the same time as
the accretion into the sink particle ends (Figure \ref{N6}).  
Because the disc mass is lower, the spin of the inner disc is perturbed 
more strongly by incoming material and so the stellar and disc 
spins are generally less aligned with each other than in the low-viscosity case.
At the end of the computation, the stellar mass is $M_\ast=0.122~{\rm M}_\odot$,
the disc mass is only $M_{\rm d} = 3.0\times 10^{-5}~{\rm M}_\odot = 2.5 \times 10^{-4}~M_\ast$, 
the accretion rate on to the star is only
$\dot{M} \approx 2.3 \times 10^{-9}~{\rm M}_\odot$
yr$^{-1}$, and the angle between the stellar and disc rotation axes is
$21.3^\circ$.

\subsection{With sink particle interaction}

We have seen in Figure \ref{N1} that thoughout the time accretion is
taking place there is a nearby sink particle in close attendance. For
the first $2 \times 10^4$ yr or so the sink particle is sufficiently
far away ($R_{\rm next} \approx 20$ au) that it does not interfere
strongly with the evolution of the disc, although its occasional
sorties to within 10 au, imply (under our assumptions of disc
truncation) that loss of angular momentum from the outer disc regions
can be significant. For the final $\approx 5000$ yr of accretion, the
neighbouring sink particle regularly approaches within 4 au, before
becoming unbound. Thus during late accretion the disc is severely
truncated. This implies that the material which is in the disc at the
end of the computation will have arrived only shortly before the
neighbouring sink particle vanished and accretion terminated. For
comparison with the above we take $\alpha_1 = 0.02$. In Figure \ref{N3} we
show the amount of material which has accreted onto the central star as
a function of time (long-dashed blue line). 
At the end of the computation the stellar mass is
$M_\ast = 1.1 \times 10^{-2}~{\rm M}_\odot$ and the disc mass is $M_{\rm d}
= 2.2 \times 10^{-4}~{\rm M}_\odot = 2.1 \times 10^{-3} M_\ast$. The
accretion rate is $\dot{M} \approx 10^{-10}~{\rm M}_\odot$ yr$^{-1}$. Thus
the effect of a nearby companion has been to severely decrease the
amount of mass accreted onto the central star. This is, of course, a
direct result of our somewhat extreme assumptions about
truncation.

In Figure \ref{N4} we show the angular coordinates of the stellar angular
momentum vector $(\theta_\ast, \phi_\ast)$ and of the inner disc spin
vector $(\theta_{\rm d}, \phi_{\rm d})$ as functions of time. In this
case it is apparent that while the companion sink particle keeps its
distance, the spin axes evolve more or less in tandem as before,
except that since the disc is now less massive it is more easily
buffeted by incoming material. Thus we see that the disc spin axis can
be quite variable throughout the accretion process (The points here
are separated by 10 year intervals.). Moreover, during the brief final
episode of chaotic close encounters disc spin axis becomes erratic as
the disc is stripped.  The upshot of this is that the final disc spin
axis is unrelated to that of the star, with the angular distance
between them being 122$^\circ$.

\section{Discussion}

We have considered in an approximate and idealised manner what happens
to gas when it falls into a sink particle in a numerical simulation of
a cluster of stars. Material is accreted onto the sink particles at a
variable rate and from a variety of directions. The aim of our
calculation has been to try to understand how the direction of the
stellar rotation axis might evolve, how the direction of the spin axis
of the inner disc (presumably related to the direction of any
accretion-induced jet) might evolve, and to what extent, if any, the
final stellar spin direction and the final disc plane (and hence any
resulting planetary orbits) might be related.

We have limited ourselves to considering the evolution of a single
sink particle from a particular simulation. In addition because of the
limitations of the data available, and the nature of assumed accretion
by sink particles, the nature of the computations we are able to carry
out are of necessity not fully self-consistent and contain various
idealisations. Thus our efforts here should be regarded as an
illustrative exploration of the kind of things that might happen
rather than definitive predictions. Nevertheless we feel that are able
to draw some general conclusions which indicate that further
consideration of this problem will be worthwhile.

The evolution of the direction of the stellar spin axis is probably
reasonably well modelled. It illustrates how the spin axis might
evolve given that the angular momentum of the material from which the
star is forming is significantly variable in direction.  The total
stellar spin direction converges quite quickly to its final
value. This is because the inner disc direction is controlled quite
closely by the plane of the outer disc (equation~\ref{nu2time}) and
thus the average spin direction of material arriving in the disc is
communicated efficiently to the star. Thus to a first approximation
the stellar spin direction just reflects the sum of everything that
arrives and once most of the mass has been accreted it is essentially
fixed. What we do not know, however, is how the direction of the
surface spin of the star relates to its total angular momentum. Since
these stars are fully convective for much of the time it may be that
the surface follows the total quite well. But we have here a
significant difference from the usual considerations in that the
accretion does not all occur axi-symmetrically and thus the internal
mixing of angular momentum in the star is much more complicated than
what is usually considered. This has implications for dynamo driving
-- the driving is likely to be quite severe as the star sorts out the
fact that different shells started rotating about different axes. Thus
the usual relationships between surface rotation and magnetic activity
might need to be modified. Thus for very young stars we might expect
that the dynamo activity depends as much on accretion history as on
its surface rotation rate. Further, as stars with masses above
0.4~${\rm M}_\odot$ approach the main sequence, their cores
become radiative and so able to decouple from the convective outer
layers. It might be that these central radiative regions can retain some
memory of the variable spin direction of the formation process. 

As a measure of the direction of disc spin we have taken the direction
of the angular momentum vector of the innermost disc radii. We do this
for two reasons. First, while accretion is active, this is likely the
direction of any accretion-driven jet. And, second, because the warp
smoothing timescale (equation~\ref{nu2time}) is relatively short, once
accretion onto the disc has ceased this is a good measure of the spin
of the final disc as a whole. Given the limitations of the analysis
discussed above, we limit ourselves to general comments about our
expectations for the evolution of the direction of disc spin. We take
as two possibly extreme examples of the nature of the accretion
process the case when the disc retains a substantial fraction of the
total mass (Figures \ref{N1}, \ref{N2}) and the case when 
the disc contains little mass, either because it is severely truncated by 
interactions with a neighbouring star (Figures \ref{N3}, \ref{N4}) or 
has undergone rapid accretion (Figures \ref{N5}, \ref{N6}).

Since the accretion process is variable both with regard to mass flux
and with regard to vectorial angular momentum direction, the disc spin
can in both cases be variable during the process of star
formation. While the angular momentum of the accreted material is
relatively high, so that the mass is added at relatively large radii,
the accretion process is smoothed by the diffusive effects of the disc
and the direction changes fairly smoothly.  However, as is
particularly evident in the truncated case for which the disc mass is
low, there are times when the angular momentum of the material being
accreted is sufficiently low that material is added at small radii and
the disc direction changes rapidly. Thus during the period of close
interaction when the disc is continuously truncated by repeated close
encounters, the disc mass is low and so the smoothing effect of the
disc is reduced. At these times the disc spin direction can vary quite
strongly and quite rapidly. This also occurs without truncation if the disc
mass is low because the viscosity is high.  It has long been recognised that the
outflows from young stars -- the Herbig-Haro flows and jets -- can
provide probes of early stellar evolution (Reipurth \& Bally,
2001). It was argued early on (Stahler 1994) that the broad molecular
flows seen around young objects are most likely driven by entrainment
by high speed jets emanating close to the central star and it was also
clear that a jet with variable direction is more able to entrain
material efficiently (Stone \& Norman, 1994; Smith et al., 1997). 
In addition there is considerable evidence that
jets and collimated outflows vary both in outflow rate and direction
(see for example Cunningham, Moeckel \& Bally, 2009; and the reviews
by Bally, Reipurth \& Davis, 2007; Bally, 2007). In addition we have
seen that accretion of mass and angular momentum becomes particularly
erratic during close interactions with companions, in line with the
ideas of Reipurth (2000).

We have seen that once accretion onto the disc has ceased, the disc
settles down to its final planar configuration well before the disc is
able to drain onto the star. This occurs because typically $\nu_2 \gg
\nu_1$. The correspondence between the spin axis of the final disc and
the spin axis of the star depends on the extent to which the final
disc has memory of the direction vector of the total anglar momentum
accreted onto the disc, and thence communicated to the star. Thus if
the disc remains essentially undisturbed throughout accretion process,
the stellar and disc rotation axes are reasonably well
aligned. However, if the outer parts of the disc (where most of the
angular momentum resides) are lost, then the final disc direction
corresponds to the spin axis of the most recently accreted
material. This axis might or might not be related to the spin of the
star.  Similarly, the action of removing the disc itself (i.e. a dynamical
encounter with another object) may misalign the remnant disc
from the stellar spin axis.

It is interesting to speculate about the possible dependence of
misalignment on stellar mass.  For example, in the hydrodynamical
calculation presented by Bate et al.\ (2002, 2003) and subsequent 
calculations brown dwarfs are produced when protostars that have
recently begun to form suffer dynamical encounters and are kicked
out of the cloud terminating their accretion before they have been
able to accrete to stellar masses.  In this case, the brown dwarf may be
left with a misaligned disc if either the disc and stellar rotation axes
had not had time to converge (because the object had recently formed)
or if the dynamical encounter itself changed the discs rotation axis.
Because the object has been ejected from the cloud it is unlikely to
subsequently accrete material with different angular momentum.
Conversely, intermediate and massive stars in such hydrodynamical
calculations typically suffer many dynamical encounters with other 
objects (e.g. Bate et al. 2003; Bate \& Bonnell 2005; Bonnell \& Bate
2005).  These encounters often leave the star embedded in the
dense gas and, thus, repeatedly destroy their discs but also allow them
to reform their discs from freshly accreted material.  Again, this may
lead to discs being misaligned with the stellar rotation axis if the
majority of the stellar mass was accreted before the final encounter
and growth of a new disc.

From the observations of transits of hot Jupiters, recent measurements 
of the angle between the planetary orbital plane and the stellar rotation 
direction show that in a substantial fraction of cases the orbital plane 
is misaligned with the stellar rotation axis (e.g. H\'ebrard et al., 2008;  
Winn et al. 2009a, 2009b; Pont et al., 2009a, 2009b; Gillon 2009; 
Johnson et al., 2009; Narita et al. 2009). 
Winn et al. (2009a)
conclude that a model on misalignment angles which supposes two
planetary populations -- one perfectly aligned and one isotropically
orientated -- is consistent with the data. They interpret this as being
due to two different causes of inward planetary migration -- one via a
gaseous disc and one via planet--planet scattering. We have found,
however, that even gaseous discs can show substantial variation in
misalignment angles depending on the history or accretion and more
especially on the history of close stellar interactions while
accretion is still taking place.

However, it also needs to be borne in mind that the accretion of the
disc material remaining at the end of our calculations can still
affect the stellar rotation. A number of authors (Cameron \& Campbell,
1993; Yi, 1994; Ghosh, 1995; Armitage \& Clarke, 1996) have argued
that the rotation rates of young stars can be regulated by magnetic
linkage between the star and a surrounding accretion disc. By the same
argument, the misalignment of spins between disc and star could also
be regulated by the same mechanism. For example the amount of mass
$\Delta M$ needed to significantly change the stellar spin is of order
$\Delta M/M_\ast \sim (k R_\ast/R_M)^2$ where $k R_\ast$ is the star's
radius of gyration and $R_M$ is the magnetospheric radius. For typical
values of $k^2 \approx 0.2$ (Armitage \& Clarke, 1996) and $R_M/R_\ast
\approx$ 5--10 (Gregory et al., 2008) we estimate $\Delta M/M_\ast
\sim 10^{-2} - 10^{-3}$. We note further that formation of the solar
planets requires a 'minimum solar nebular' mass of around $M_{\rm
  d}/M_\ast \ge 0.02$. Determining how much of this mass is actually
able to interact with the central star and so change its spin will
depend on the details of the planet formation process, and in
particular on the timing and extent of the formation of any inner gap
which would decouple disc and star.

\section{Conclusions}

We have considered the evolution of the stellar and disc spin axes
during the formation of a star which is accreting in a variable
fashion from an inherently chaotic environment. To model this process
we have modelled the evolution of a warped disc which has material
added to it in a variable fashion. We take the input of mass and
angular momentum to the disc to that acquired by the 'sink particle'
in an SPH simulation of the formation of a cluster of stars in from
self-gravitating turbulent gas. We have noted that many of the
assumptions we have used cannot be made self-consistent. Thus we
caution against drawing specific conclusions from our analysis. The
calculations we show here should be regarded simply as
illustrative. Nevertheless there are some general points
which can be made. 

First, the variability of the direction of the spin
of material accreting onto the central protostar implies that the
internal velocity field of such stars may be more complicated than the
usual assumption of aligned differential rotation. 

Second the lighter
the disc (and the disc can lose mass either through rapid accretion or
through interaction with a nearby protostar) the more able is the
accreted material to cause the inner disc spin, which we identify with
the direction of a jet, to vary. 

Third, the final stellar rotation axis and the
final disc spin axis can be strongly misaligned. However, this occurs
most strongly when the disc is truncated so that rotation direction of
the final disc material depends simply on what fell in last. In this
case, although the disc might be misaligned with the star, it might
not contain enough material to form planets. And we have noted that,
depending on the details of the models, it may be that a misaligned
disc which is massive enough to form planets, may also be massive
enough to reduce the misalignment.

In conclusion, we have shown that it should be possible make some
deductions about the accretion history of a young star by observations
of jet direction variability, of star/planetary orbit misalignments,
and perhaps even of its internal rotation structure. It is clear that
further work is required to improve models of the star formation
process with better account being taken of the nature of the accretion
process within the final 100 au or so.

\section*{Acknowledgments} 

The authors thank Rebecca Martin for correcting an error 
in the code and John Bally 
for communicating material prior to publication. 
MRB is grateful for the support of a EURYI Award.  
This work, conducted as part of the award ÒThe formation of stars and 
planets: Radiation hydrodynamical and magnetohydrodynamical 
simulationsÓ made under the European Heads of Research 
Councils and European Science Foundation EURYI 
(European Young Investigator) Awards scheme, was supported 
by funds from the Participating Organisations of EURYI and the 
EC Sixth Framework Programme.

\section*{Appendix: Addition and shifting of mass}

We discuss here our algorithm for adding material to the disc.

Suppose that the disc grid cells are at radii $R_i, i=1,N$
corresponding to specific angular momenta $h_i, i=1,N$. In a timesetep
$\Delta t$ we add a mass $\Delta M$ with angular momentum ${\bf \Delta
  J}$. Thus the specific angular momentum of the added material is
\begin{equation}
h = \frac{\Delta J}{\Delta M},
\end{equation}
where $\Delta J = | {\bf \Delta J} |$. Matter is than added to the
cells $k$ and $k+1$ where $h_k < h < h_{k+1}$.   

Then the angular momentum in cell $k$ is incremented by an amount
\begin{equation}
{\bf \Delta J}_k = \frac{h_{k+1} - h}{h_{k+1} - k_k} {\bf \Delta J} ,
\end{equation}
and that in cell $k+1$ by an amount
\begin{equation}
{\bf \Delta J}_{k+1} = \frac{h - h_k}{h_{k+1} - k_k} {\bf \Delta J} .
\end{equation}
In this manner angular momentum is exactly conserved. 

However, the amount of mass actually added is now
\begin{equation}
\Delta M^\prime = \frac{\Delta J_{k+1}}{h_{k+1}} + \frac{\Delta
  J_k}{h_k},
\end{equation}
which needs to be compared with the actual mass to be added $\Delta M
= \Delta J/h$.

If we write $h_{k+1} = h(1 + \epsilon_2)$ and $h_k = h(1 -
\epsilon_1)$, where $\epsilon_2 > 0$ and $\epsilon_1 > 0$ then to first
order in the small quantities $\epsilon$ it is simple to show that
\begin{equation}
\frac{\Delta M^\prime - \Delta M}{\Delta M} = \epsilon_2 \, \epsilon_1 >
0.
\end{equation}

Thus the algorithm for the addition of angular momentum leads to a
systematic over-estimate of the amount of mass added of an amount
which is on average (i.e. when $\epsilon_1$ and $\epsilon_2$ are of
comparable magnitude) second order in $dR/R$.

\label{lastpage}

\end{document}